\documentclass[11pt]{article}

\usepackage{url}
\usepackage[breaklinks,colorlinks,bookmarks]{hyperref}

\usepackage{fullpage}
\usepackage{latexsym}
\usepackage{amssymb}
\usepackage{amsfonts}

\def\01{\{0,1\}}

\newcommand{\Oh}[1]{\mathrm{O}\mathchoice{\!}{\!}{}{}\left(#1\right)}
\newcommand{\OO}[1]{\mathrm{O}(#1)}
\newcommand{\oo}[1]{\mathrm{o}(#1)}
\newcommand{\Om}[1]{\Omega\mathchoice{\!}{\!}{}{}\left(#1\right)}
\newcommand{\Th}[1]{\Theta\mathchoice{\!}{\!}{}{}\left(#1\right)}
\newcommand{\eps}{\varepsilon}
\newcommand{\OR}{\mbox{\rm OR}}
\newcommand{\Search}{\mbox{\rm Search}}

\newcommand{\DISJ}{\mbox{\rm DISJ}}
\newcommand{\poly}{\mbox{\rm poly}}

\newcommand{\ket}[1]{|#1\rangle}

\newcommand{\norm}[1]{\mbox{$\parallel{#1}\parallel$}}
\newcommand{\inpc}[2]{\langle{#1},{#2}\rangle} 
\newcommand{\Tr}{\mbox{\rm Tr}}
\newcommand{\Adv}[1]{\mathrm{Adv}(#1)}
\newcommand{\suc}[2]{\mathrm{Suc}_{#1}(#2)}

\newcommand{\parity}[2]{{#1^{\oplus #2}}}
\newcommand{\parityof}[2]{#1 \oplus \dots \oplus #2}
\newcommand{\vectorres}[2]{{#1^{(#2)}}}
\newcommand{\bs}{\mathit{bs}}
\renewcommand{\Pr}[1]{\mathrm{Pr}\left[ #1 \right]}

\newtheorem{definition}{Definition}
\newtheorem{theorem}{Theorem}
\newtheorem{lemma}[theorem]{Lemma}
\newtheorem{corollary}[theorem]{Corollary}
\newenvironment{remark}[1][Remark.]{
	\par\medskip
	\noindent {\em #1}
}{
	\par\medskip
}

\newenvironment{proof}[1][Proof.]{
	\par
	\noindent \textbf{#1}
}{
	\nobreak\leavevmode
	\hfill $\Box$\par\bigskip
}

\begin{document}

\title{Quantum and Classical Strong Direct Product Theorems\\ and Optimal Time-Space Tradeoffs}
\author{Hartmut Klauck\thanks{Supported by Canada's NSERC and MITACS, and by DFG grant KL 1470/1.}\\
University of Calgary\\
klauckh@cpsc.ucalgary.ca
 \and
Robert \v Spalek\thanks{Supported in part by the EU fifth framework project RESQ, IST-2001-37559.}\\
CWI, Amsterdam\\
sr@cwi.nl
\and
\addtocounter{footnote}{-1}
Ronald de Wolf\footnotemark\\
CWI, Amsterdam\\
rdewolf@cwi.nl}
\date{}
\maketitle

\begin{abstract}
A strong direct product theorem says that if we want to compute
$k$ independent instances of a function, using less than $k$ times
the resources needed for one instance, then our overall success
probability will be exponentially small in $k$.
We establish such theorems for the classical as well as quantum
query complexity of the \OR\ function. This implies slightly
weaker direct product results for all total functions.
We prove a similar result for quantum communication
protocols computing $k$ instances of the Disjointness function.

Our direct product theorems imply a time-space tradeoff
$T^2S=\Om{N^3}$ for sorting $N$ items on a quantum computer, which
is optimal up to polylog factors. They also give several tight
time-space and communication-space tradeoffs for the problems of
Boolean matrix-vector multiplication and matrix multiplication.
\end{abstract}

\section{Introduction}

\subsection{Direct product theorems}

For every reasonable model of computation one can ask the following
fundamental question:
\begin{quote}
How do the resources that we need for computing $k$ independent
instances of $f$ scale with the resources needed for one instance and with $k$?
\end{quote}
Here the notion of ``resource'' needs to be specified.  It could refer to
time, space, queries, communication etc. Similarly we need to
define what we mean by ``computing $f$'', for instance whether we
allow the algorithm some probability of error, and whether this
probability of error is average-case or worst-case.

In this paper we consider two kinds of resources, queries and
communication, and allow our algorithms some error probability.
An algorithm is given $k$ inputs $x^1,\ldots,x^k$, and has
to output the vector of $k$ answers $f(x^1),\ldots,f(x^k)$.
The issue is how the algorithm can optimally distribute its
resources among the $k$ instances it needs to compute.
We focus on the relation between the total amount $T$
of resources available and the best-achievable success
probability $\sigma$ (which could be average-case or worst-case).
Intuitively, if every algorithm with $t$
resources must have some constant error probability when
computing \emph{one} instance of $f$, then for computing $k$ instances
we expect a constant error on each instance and hence an
exponentially small success probability for the $k$-vector as a whole.
Such a statement is known as a \emph{weak} direct product theorem:
\begin{quote}
If $T\approx t$, then $\sigma=2^{-\Om k}$
\end{quote}
However, even if we give our algorithm roughly $k t$ resources,
on average it still has only $t$ resources available per instance.
So even here we expect a constant error per instance and
an exponentially small success probability overall.
Such a statement is known as a \emph{strong} direct product theorem:
\begin{quote}
If $T\approx k t$, then $\sigma=2^{-\Om k}$
\end{quote}
Strong direct product theorems, though intuitively very plausible,
are generally hard to prove and sometimes not even true.
Shaltiel~\cite{shaltiel:sdpt} exhibits a general class of
examples where strong direct product theorems fail.
This applies for instance to query complexity, communication complexity,
and circuit complexity.
In his examples, success probability is taken under the uniform
probability distribution on inputs. The function is
chosen such that for most inputs, most of the $k$ instances
can be computed quickly and without any error probability.
This leaves enough resources to solve the few hard instances
with high success probability.  Hence for his functions,
with $T\approx t k$, one can achieve average success
probability close to 1.

Accordingly, we can only establish direct product theorems in
special cases. Examples are Nisan et al.'s~\cite{nrs:products}
strong direct product theorem for ``decision forests'', Parnafes
et al.'s~\cite{prw:productgcd} direct product theorem for
``forests'' of communication protocols, Shaltiel's strong direct
product theorems for ``fair'' decision trees and his discrepancy
bound for communication complexity~\cite{shaltiel:sdpt}. In the
quantum case, Aaronson~\cite[Theorem~10]{aaronson:advicecomm}
established a result for the unordered search problem that lies
in between the weak and the strong theorems: every $T$-query
quantum algorithm for searching $k$ marked items among $N=k n$
input bits will have success probability $\sigma\leq \Oh{T^2/N}^k$.
In particular, if $T\ll\sqrt{k n}$, then $\sigma=2^{-\Om{k}}$.

Our main contributions in this paper are strong direct product
theorems for the $\OR$-function in various settings.
First consider the case of classical randomized algorithms.
Let $\OR_n$ denote the $n$-bit $\OR$-function, and
let $\vectorres f k$ denote $k$ independent instances of a function $f$.
Any randomized algorithm with less than, say, $n/2$ queries
will have a constant error probability when computing $\OR_n$.
Hence we expect an exponentially small success probability when
computing $\vectorres {\OR_n} k$ using $\ll k n$ queries.
We prove this in Section~\ref{secclasdpt}:
\begin{quote}
{\bf SDPT for classical query complexity:}\\
Every randomized algorithm that computes $\vectorres {\OR_n} k$ using
$T\leq \alpha k n$ queries has worst-case success probability $\sigma=2^{-\Om k}$
(for $\alpha>0$ a sufficiently small constant).
\end{quote}
For simplicity we have stated this result with $\sigma$ being
\emph{worst-case} success probability, but the statement is also
valid for the \emph{average} success probability
under a hard $k$-fold product distribution that is implicit in our proof.

This statement for $\OR$ actually implies a somewhat weaker DPT for all
\emph{total} functions $f$, via the notion of \emph{block sensitivity} $\bs(f)$.
Using techniques of Nisan and Szegedy~\cite{nisan&szegedy:degree},
we can embed $\OR_{\bs(f)}$ in $f$ (with the promise that the weight
of the \OR's input is 0 or 1), while on the other hand we know
that the classical bounded-error query complexity $R_2(f)$
is upper bounded by $\bs(f)^3$~\cite{bbcmw:polynomialsj}.
This implies:
\begin{quote}
Every randomized algorithm that computes $\vectorres f k$ using
$T\leq \alpha k R_2(f)^{1/3}$ queries has worst-case success probability
$\sigma=2^{-\Om k}$.
\end{quote}
This theorem falls short of a true strong direct product theorem
in having $R_2^{1/3}(f)$ instead of $R_2(f)$ in the resource bound.
However, the other two main aspects of a SDPT remain valid:
the linear dependence of the resources on $k$ and the exponential
decay of the success probability.

Next we turn our attention to \emph{quantum} algorithms.
Buhrman et~al.~\cite{bnrw:robustq} actually proved that roughly $k$ times
the resources for one instance suffices to compute $\vectorres f k$ with success
probability \emph{close to 1}, rather than exponentially small:
$Q_2(\vectorres f k)=\Oh{k Q_2(f)}$, where $Q_2(f)$ denotes
the quantum bounded-error query complexity of $f$
(such a result is not known to hold in the classical world).
For instance, $Q_2(\OR_n)=\Th{\sqrt{n}}$ by Grover's search algorithm,
so $\Oh{k\sqrt{n}}$ quantum
queries suffice to compute $\vectorres {\OR_n} k$ with high success probability.
In Section~\ref{secquadpt} we show that if we make the number of queries
slightly smaller, the best-achievable success probability suddenly
becomes exponentially small:
\begin{quote}
{\bf SDPT for quantum query complexity:}\\
Every quantum algorithm that computes $\vectorres {\OR_n} k$ using
$T\leq \alpha k \sqrt{n}$ queries has worst-case success probability
$\sigma=2^{-\Om{k}}$ (for $\alpha>0$ a sufficiently small constant).
\end{quote}
Our proof uses the polynomial method~\cite{bbcmw:polynomialsj}
and is completely different from the classical proof.
The polynomial method was also used by
Aaronson~\cite{aaronson:advicecomm} in his proof of a weaker quantum
direct product theorem for the search problem, mentioned above.
Our proof takes its starting point from his proof, analyzing the
degree of a single-variate polynomial that is 0 on $\{0,\ldots,k-1\}$,
at least $\sigma$ on $k$, and between 0 and 1 on $\{0,\ldots,k n\}$.
The difference between his proof and ours is that we partially factor
this polynomial, which gives us some nice extra properties over
Aaronson's approach of differentiating the polynomial, and we use a strong
result of Coppersmith and Rivlin~\cite{coppersmith&rivlin:poly}.
In both cases (different) extremal properties of
Chebyshev polynomials finish the proofs.

Again, using block sensitivity we can obtain a weaker result for
all total functions:
\begin{quote}
Every quantum algorithm that computes $\vectorres f k$ using $T\leq \alpha k
Q_2(f)^{1/6}$
queries has worst-case success probability $\sigma=2^{-\Om{k}}$.
\end{quote}
The third and last setting where we establish a strong direct product theorem
is quantum communication complexity.
Suppose Alice has an $n$-bit input $x$ and Bob has an $n$-bit input $y$.
These $x$ and $y$ represent sets, and $\DISJ_n(x,y)=1$ iff those sets are disjoint.
Note that $\DISJ_n$ is the negation of $\OR_n(x\wedge y)$, where
$x\wedge y$ is the $n$-bit string obtained by bitwise AND-ing $x$ and $y$.
In many ways, $\DISJ_n$ has the same central role in communication
complexity as $\OR_n$ has in query complexity.
In particular, it is ``co-NP complete''~\cite{bfs:classes}.
The communication complexity of $\DISJ_n$ has been well studied:
it takes $\Th n$ bits of communication
in the classical world~\cite{ks:disj,razborov:disj}
and $\Th{\sqrt{n}}$ in the quantum world
\cite{BuhrmanCleveWigderson98,hoyer&wolf:disjeq,aaronson&ambainis:search,razborov:qdisj}.
For the case where Alice and Bob want to compute $k$ instances
of Disjointness, we establish a strong direct product theorem in Section~\ref{secquacomdpt}:
\begin{quote}
{\bf SDPT for quantum communication complexity:}\\
Every quantum protocol that computes $\vectorres {\DISJ_n} k$ using
$T\leq \alpha k \sqrt{n}$ qubits of communication
has worst-case success probability $\sigma=2^{-\Om{k}}$.
\end{quote}
Our proof uses Razborov's~\cite{razborov:qdisj} lower bound technique
to translate the quantum protocol to a polynomial, at which point the
polynomial results established for the quantum query SDPT take over.
We can obtain similar results for other symmetric predicates.

One may also consider algorithms that compute the \emph{parity}
of the $k$ outcomes instead of the vector of $k$ outcomes.
This issue has been well studied, particularly in circuit complexity, and
generally goes under the name of \emph{XOR lemmas}~\cite{yao:xor,gnw:xor}.
In this paper we focus mostly on the vector version, but we can prove
similar strong bounds for the parity version. In particular,
we state a classical strong XOR lemma in Section~\ref{secclassicalparity}
and can get similar strong XOR lemmas for the quantum case
using the technique of Cleve et~al.~\cite[Section~3]{cdnt:ip}.
They show how the ability to compute the parity of any subset of $k$
bits with probability $1/2+\eps$, suffices to compute the full
$k$-vector with probability $4\eps^2$.
Hence our strong quantum direct product theorems imply strong quantum XOR lemmas.

\subsection{Time-Space and Communication-Space tradeoffs}

Apart from answering a fundamental question about the
computational models of (quantum) query complexity and
communication complexity, our direct product theorems also imply a
number of new and optimal time-space tradeoffs.

First, we consider the tradeoff between the time $T$ and space $S$
that a quantum circuit needs for \emph{sorting} $N$ numbers.
Classically, it is well known that $TS=\Om{N^2}$ and that this tradeoff is
achievable~\cite{beame:tradeoff}. In the quantum case,
Klauck~\cite{klauck:qsorting} constructed a bounded-error quantum
algorithm that runs in time $T=\OO{(N\log
N)^{3/2}/\sqrt{S}}$ for all $(\log N)^3\leq S\leq N/\log N$. He
also showed\footnote{Unfortunately there is an error in the proof
presented in \cite{klauck:qsorting}, namely Lemma 5 appears to be
wrong.} a lower bound $TS=\Om{N^{3/2}}$, which is close to optimal
for small $S$ but not for large $S$. We use our strong direct
product theorem to establish the tradeoff $T^2S=\Om{N^3}$. This is
tight up to polylogarithmic factors.

Secondly, we consider time-space and communication-space tradeoffs
for the problems of \emph{Boolean matrix-vector product} and \emph{Boolean
matrix product}. In the first problem there are an $N\times N$
matrix $A$ and a vector $b$ of dimension $N$, and the goal is to
compute the vector $c=Ab$, where $c_i=\vee_{j=1}^N \left(A[i,j]\wedge
b_j\right)$. In the setting of time-space tradeoffs, the matrix $A$ is
fixed and the input is the vector $b$. In the problem of matrix
multiplication two matrices have to be multiplied with the same
type of Boolean product, and both are inputs.

Time-space tradeoffs for Boolean matrix-vector multiplication have
been analyzed in an average-case scenario by Abrahamson
\cite{abrahamson:booleantrade}, whose results give a worst-case
lower bound of $TS=\Om{N^{3/2}}$ for classical algorithms. He
conjectured that a worst-case lower bound of $TS=\Om{N^2}$
holds. Using our classical direct product result we are able to
confirm this, i.e., there is a matrix $A$, such that computing $Ab$
requires $TS=\Om{N^2}$. We also show a lower bound of
$T^2 S=\Om{N^{3}}$ for this problem in the quantum case.
Both bounds are tight (the second within a logarithmic factor) if
$T$ is taken to be the number of queries to the inputs. We also
get a lower bound of $T^2 S=\Om{N^5}$ for the problem
of multiplying two matrices in the quantum case.
This bound is close to optimal for small $S$; it is open whether
it is close to optimal for large $S$.

Research on communication-space tradeoffs in the classical setting
has been initiated by Lam et al.~\cite{lam:commtrade} in a
restricted setting, and by Beame et al.~\cite{beame:commtrade} in
a general model of space-bounded communication complexity. In the
setting of communication-space tradeoffs, players Alice and Bob are
modeled as space-bounded circuits, and we are interested in the
communication cost when given particular space bounds. For the
problem of computing the matrix-vector product Alice receives the
matrix $A$ (now an input) and Bob the vector $b$. Beame et
al.~gave tight lower bounds e.g.~for the matrix-vector product and
matrix product over GF(2), but stated the complexity of Boolean
matrix-vector multiplication as an open problem. Using our direct
product result for quantum communication complexity we are able to
show that any quantum protocol for this problem satisfies
$C^2 S=\Om{N^3}$. This is tight within a polylogarithmic factor.
We also get a lower bound of $C^2 S=\Om{N^5}$ for
computing the product of two matrices, which again is tight.

Note that no classical lower bounds for these problems were known
previously, and that finding better classical lower bounds than
these remains open. The possibility to show good quantum bounds
comes from the deep relation between quantum protocols and
polynomials implicit in Razborov's lower bound technique
\cite{razborov:qdisj}.

\section{Preliminaries}

\subsection{Quantum query algorithms}

We assume familiarity with quantum computing~\cite{nielsen&chuang:qc}
and sketch the model of quantum query complexity,
referring to~\cite{buhrman&wolf:dectreesurvey} for more
details, also on the close relation between query complexity
and degrees of multivariate polynomials.
Suppose we want to compute some function $f$.
For input $x\in\01^N$, a \emph{query} gives us access to the
input bits. It corresponds to the unitary transformation
$$
O:\ket{i,b,z}\mapsto\ket{i,b\oplus x_i,z}.
$$
Here $i\in[N]=\{1,\ldots,N\}$ and $b\in\01$; the $z$-part corresponds
to the workspace, which is not affected by the query.
We assume the input can be accessed only via such queries.
A $T$-query quantum algorithm has the form $A=U_T O U_{T-1}\cdots O U_1 O U_0$,
where the $U_k$ are fixed unitary transformations, independent of $x$.
This $A$ depends on $x$ via the $T$ applications of $O$.
The algorithm starts in initial $S$-qubit state $\ket{0}$ and
its \emph{output} is the result of measuring a dedicated part
of the final state $A\ket{0}$.
For a Boolean function $f$, the output
of $A$ is obtained by observing the leftmost qubit of the
final superposition $A\ket{0}$, and its
\emph{acceptance probability} on input $x$
is its probability of outputting 1.

One of the most interesting quantum query algorithms is
Grover's search algorithm~\cite{grover:search,bbht:bounds}.
It can find an index of a 1-bit in an $n$-bit input in
expected number of $\Oh{\sqrt{n/(|x|+1)}}$ queries,
where $|x|$ is the Hamming weight (number of ones) in the input.
If we know that $|x|\leq 1$, we can solve the search problem exactly
using $\frac{\pi}{4}\sqrt{n}$ queries~\cite{bhmt:countingj}.

For investigating time-space tradeoffs we use the circuit model. A
circuit accesses its input via an oracle like a query algorithm.
Time corresponds to the number of gates in the circuit. We will,
however, usually consider the number of queries to the input,
which is obviously a lower bound on time. A quantum circuit uses
space $S$ if it works with $S$ qubits only. We require that the
outputs are made at predefined gates in the circuit, by writing
their value to some extra qubits that may not be used later on.
Similar definitions are made for classical circuits.

\subsection{Communicating quantum circuits}

In the model of quantum communication complexity, two players Alice
and Bob compute a function $f$ on distributed inputs $x$ and $y$.
The complexity measure of interest in this setting is the amount of
communication. The players follow some predefined protocol that
consists of local unitary operations, and the exchange of qubits.
The communication cost of a protocol is the maximal number of
qubits exchanged for any input.
In the standard model of communication complexity, Alice and Bob
are computationally unbounded entities, but we are also interested
in what happens if they have bounded memory, i.e., they work with
a bounded number of qubits. To this end we model Alice and Bob as
communicating quantum circuits, following Yao~\cite{yao:qcircuit}.

A pair of communicating quantum circuits is actually a single
quantum circuit partitioned into two parts. The allowed operations
are local unitary operations and access to the inputs that are
given by oracles. Alice's part of the circuit may use oracle gates
to read single bits from her input, and Bob's part of the circuit
may do so for his input. The communication $C$ between the two parties
is simply the number of wires carrying qubits that cross between
the two parts of the circuit.
A pair of communicating quantum circuits uses space $S$, if the
whole circuit works on $S$ qubits.

In the problems we consider, the number of outputs is much larger
than the memory of the players. Therefore we use the following output
convention. The player who computes the value of an output sends
this value to the other player at a predetermined point in the
protocol.
In order to make the models as general as possible, we furthermore
allow the players to do local measurements, and to throw qubits
away as well as pick up some fresh qubits. The space requirement
only demands that at any given time no more than $S$ qubits are in
use in the whole circuit.

A final comment regarding upper bounds: Buhrman
et~al.~\cite{BuhrmanCleveWigderson98} showed how to run
a query algorithm in a distributed fashion with small
overhead in the communication.
In particular, if there is a $T$-query quantum algorithm computing
$N$-bit function $f$, then there is a pair of communicating
quantum circuits with $O(T\log N)$ communication that computes
$f(x\wedge y)$ with the same success probability.
We refer to the book of Kushilevitz and Nisan~\cite{kushilevitz&nisan:cc}
for more on communication complexity in general,
and to the surveys~\cite{klauck:qccsurvey,buhrman:qccsurvey,wolf:qccsurvey}
for more on its quantum variety.

\section{Strong Direct Product Theorem for Classical Queries}\label{secclasdpt}

In this section we prove a strong direct product theorem for classical randomized
algorithms computing $k$ independent instances of $\OR_n$.  By Yao's
principle, it is sufficient to prove it for deterministic algorithms under a
fixed hard input distribution.

\subsection{Non-adaptive algorithms}

We first establish a strong direct
product theorem for non-adaptive algorithms.
We call an algorithm \emph{non-adaptive} if, for each of the $k$ input blocks, the maximum number of
queries in that block is fixed before the first query.
Let $\suc {t,\mu} f$ be the success probability of the best algorithm for $f$
under $\mu$ that queries at most $t$ input bits.

\begin{lemma}
\label{lem:non-adaptive}
Let $f: \01^n \to \01$ and $\mu$ be an input distribution.
Every non-adaptive deterministic algorithm for $\vectorres f k$ under
$\mu^k$ with $T \le k t$ queries has success probability $\sigma \le \suc
{t,\mu} f ^k$.
\end{lemma}

\begin{proof}
The proof has two steps.  First, we prove by induction that non-adaptive
algorithms for $\vectorres f k$ under general product distribution $\mu_1
\times \dots \times \mu_k$ that spend $t_i$ queries in $x^i$ have success
probability $\le \prod_{i=1}^k \suc {t_i,\mu_i} f$.  Second, we argue that,
when $\mu_i = \mu$, the value is maximal for $t_i = t$.

Following~\cite[Lemma 7]{shaltiel:sdpt}, we prove the first part by induction
on $T = t_1 + \dots + t_k$.  If $T = 0$, then the algorithm
has to guess $k$ independent random variables $x^i \sim \mu_i$.  The
probability of success is equal to the product of the individual success
probabilities, i.e.  $\prod_{i=1}^k \suc {0,\mu_i} f$.

For the induction step $T\Rightarrow T+1$:
pick some $t_i \ne 0$ and consider two input distributions
$\mu'_{i,0}$ and $\mu'_{i,1}$ obtained from $\mu_i$ by fixing the queried bit
$x^i_j$.  By the induction hypothesis, for each value $b \in \01$, there is
an optimal non-adaptive algorithm $A_b$ that achieves the success probability
$\suc {t_i-1, \mu'_{i,b}} f \cdot \prod_{j\ne i} \suc {t_j, \mu_j} f$.  We
construct a new algorithm $A$ that calls $A_b$ as a subroutine after it has
queried $x^i_j$ with $b$ as an outcome.  $A$ is optimal and it has success
probability
\[
\left( \sum_{b=0}^1 \mathrm{Pr}_{\mu_i}[x^i_j=b] \cdot \suc {t_i-1, \mu'_{i,b}} f \right)
  \cdot \prod_{j \ne i} \suc {t_j, \mu_j} f
= \prod_{i=1}^k \suc{t_i, \mu_i} f.
\]

For symmetry reasons, if all $k$ instances $x^i$ are independent and identically
distributed, then the optimal distribution of queries $t_1 + \dots + t_k =
k t$ is uniform, i.e. $t_i = t$.  In such a
case, the algorithm achieves the success probability $\suc {t,\mu} f^k$.
\end{proof}

\subsection{Adaptive algorithms}

In this section we prove a similar statement also for adaptive algorithms.

\begin{remark}
The strong direct product theorem is not always true for adaptive algorithms.
Following~\cite{shaltiel:sdpt}, define $h(x) = x_1 \vee (\parityof {x_2}
{x_n})$.  Clearly $\suc {\frac23 n,\mu} h = 3/4$ for $\mu$ uniform.
By a Chernoff bound, $\suc {\frac23 n k,\mu^k} {\vectorres h k} =
1 - 2^{-\Om k}$, because approximately half of the blocks can be
solved using just 1 query and the unused queries can be used to
answer exactly also the other half of the blocks.
\end{remark}

However, the strong direct product theorem is valid for $\vectorres{\OR_n}k$ under $\nu^k$, where
$\nu(0^n)=1/2$ and $\nu(e_i)=1/2n$ for $e_i$ an $n$-bit string that contains a
1 only at the $i$-th position.
It is simple to prove that $\suc {\alpha n,\nu} {\OR_n} = {\alpha + 1 \over 2}$.
Non-adaptive algorithms for $\vectorres {\OR_n} k$ under $\nu^k$ with $\alpha
k n$ queries thus have $\sigma \le ({\alpha + 1 \over 2})^k =
2^{-\log({2 \over \alpha + 1}) k}$.  We can achieve any
$\gamma<1$ by choosing $\alpha$ sufficiently small.
We prove that adaptive algorithms cannot be much better.
Without loss of generality, we assume:
\begin{enumerate}
\item The adaptive algorithm is deterministic.
By Yao's principle~\cite{yao:unified}, if there
exists a randomized algorithm with success probability $\sigma$ under some input
distribution, then there exists a deterministic algorithm with success probability
$\sigma$ under that distribution.

\item Whenever the algorithm finds a 1 in some input block, it stops querying that block.

\item The algorithm spends the same number of queries in all blocks where it
does not find a 1.  This is optimal due to the symmetry between the blocks,
and implies that the algorithm spends at least as many queries in each
``empty'' input block as in each ``non-empty'' block.
\end{enumerate}

\begin{lemma}
\label{lem:triple}
If there is an adaptive $T$-query algorithm $A$ computing $\vectorres{\OR_n}{k}$ under
$\nu^k$ with success probability $\sigma$, then there is a non-adaptive $3T$-query
algorithm $A'$ computing $\vectorres{\OR_n}{k}$ with success probability $\sigma - 2^{-\Om k}$.
\end{lemma}

\begin{proof}
Let $Z$ be the number of empty blocks.  ${\rm E}[Z] = k/2$ and, by a Chernoff
bound, $\delta = \Pr{ Z < k/3} = 2^{-\Om k}$.  If $Z \ge k/3$,
then $A$ spends at most $3T/k$ queries in each empty block.  Define
non-adaptive $A'$ that spends $3T/k$ queries in {\em each\/} block.
Then $A'$ queries all the positions that $A$ queries, and maybe some more.
Compare the overall success probabilities of $A$ and $A'$:
\[
\begin{array}{r @{\ } l}
\sigma_A =& \Pr{ Z < k/3 } \cdot \Pr{ \mbox{$A$ succeeds} \mid Z < k/3}\\
 +& \Pr{ Z \ge k/3 } \cdot \Pr{ \mbox{$A$ succeeds} \mid Z \ge k/3} \\
\le& \delta \cdot 1
  + \Pr{Z \ge k/3} \cdot \Pr{ \mbox{$A'$ succeeds} \mid Z \ge k/3}\\
\le& \delta + \sigma_{A'}.
\end{array}
\]
We conclude that $\sigma_{A'} \ge \sigma_A - \delta$.
({\em Remark.} By replacing the $k/3$-bound on $Z$ by a $\beta k$-bound
for some $\beta > 0$, we can obtain arbitrary
$\gamma < 1$ in the exponent $\delta = 2^{-\gamma k}$, while the
number of queries of $A'$ becomes $T/\beta$.)
\end{proof}

Combining the two lemmas establishes the following theorem:

\begin{theorem}[SDPT for \OR]\label{th:classor}
For every $0<\gamma<1$, there exists an $\alpha > 0$ such that every
randomized algorithm for $\vectorres {\OR_n} k$ with $T \le \alpha k n$
queries has success probability $\sigma \le 2^{-\gamma k}$.
\end{theorem}

\subsection{A bound for the parity instead of the vector of results}\label{secclassicalparity}

Here we give a strong direct product theorem for the \emph{parity}
of $k$ independent instances of $\OR_n$.  The parity is a Boolean
variable, hence we can always guess it with probability at least $\frac12$.
However, we prove that the advantage (instead of the success probability) of
our guess must be exponentially small.

Let $X$ be a random bit with $\Pr{X=1} = p$.  We define the {\em advantage of
$X$} by $\Adv X = |2p-1|$.  Note that a uniformly distributed random bit has
advantage $0$ and a bit known with certainty has advantage $1$.  It is well
known that if $X_1, \dots, X_k$ are independent random bits, then $\Adv
{\parityof {X_1} {X_k}} = \prod_{i=1}^k \Adv{X_i}$.  Compare this with the
fact that the probability of guessing correctly the complete vector $(X_1, \dots,
X_k)$ is the product of the individual probabilities.

We have proved a lower bound for the computation of $\vectorres {\OR_n} k$
(vector of \OR's).
By the same technique, replacing the success probability by the advantage
in all claims and proofs, we can also prove a lower bound for the
computation of $\parity {\OR_n} k$ (parity of \OR's).

\begin{theorem}[SDPT for parity of \OR's]
For every $0<\gamma<1$, there exists an $\alpha > 0$ such that every
randomized algorithm for $\parity {\OR_n} k$ with $T \le \alpha k n$
queries has advantage $\tau \le 2^{-\gamma k}$.
\end{theorem}

\subsection{A bound for all functions}
\label{sec:all-func}

Here we show that the strong direct product theorem for $\OR$
actually implies a weaker direct product theorem for all functions.
In this weaker version, the success probability of computing $k$
instances still goes down exponentially with $k$, but we need to start
from a polynomially smaller bound on the overall number of queries.

\begin{definition}
For $x\in\01^n$ and $S\subseteq[n]$, we use $x^S$ to denote
the $n$-bit string obtained from $x$ by flipping the bits in $S$.
Consider a (possibly partial) function $f:\mathcal{D}\rightarrow Z$,
with $\mathcal{D}\subseteq\01^n$. The \emph{block sensitivity}
$\bs_x(f)$ of $x\in\mathcal{D}$ is the maximal $b$ for which there
are disjoint sets $S_1,\ldots,S_b$ such that $f(x)\neq f(x^{S_i})$.
The \emph{block sensitivity} of $f$ is $\max_{x\in\mathcal{D}} \bs_x(f)$.
\end{definition}

Block sensitivity is closely related to deterministic and
bounded-error classical query complexity:

\begin{theorem}[\cite{nisan:pram&dt,bbcmw:polynomialsj}]
$R_2(f)=\Om{\bs(f)}$ for all $f$,
$D(f)\le\bs(f)^3$ for all total \mbox{Boolean $f$.}
\end{theorem}

Nisan and Szegedy~\cite{nisan&szegedy:degree} showed how
to embed a $\bs(f)$-bit $\OR$-function (with the promise
that the input has weight $\leq 1$) into $f$.
Combined with our strong direct product theorem for $\OR$,
this implies a direct product theorem for all functions
in terms of their block sensitivity:

\begin{theorem}
For every $0 < \gamma < 1$, there exists an $\alpha > 0$ such that for every $f$,
every classical algorithm for $\vectorres f k$ with $T \le \alpha k \bs(f)$ queries
has success probability $\sigma \le 2^{-\gamma k}$.
\end{theorem}

This is optimal whenever $R_2(f)=\Th{\bs(f)}$, which is the
case for most functions. For total functions, the gap between
$R_2(f)$ and $\bs(f)$ is not more than cubic, hence

\begin{corollary}
For every $0<\gamma<1$, there exists an $\alpha > 0$ such that for every total
Boolean $f$, every classical algorithm for $\vectorres f k$ with $T \le \alpha k
R_2(f)^{1/3}$ queries has success probability $\sigma \le 2^{-\gamma k}$.
\end{corollary}

\section{Strong Direct Product Theorem for Quantum Queries}\label{secquadpt}

In this section we prove a strong direct product theorem for
quantum algorithms computing $k$ independent instances of $\OR$.
Our proof relies on the polynomial method of~\cite{bbcmw:polynomialsj}.

\subsection{Bounds on polynomials}\label{ssecpolynomial}

We use three results about polynomials, also used in~\cite{bcwz:qerror}.
The first is by Coppersmith and Rivlin~\cite[p.~980]{coppersmith&rivlin:poly}
and gives a general bound for polynomials bounded by 1 at integer points:

\begin{theorem}
[Coppersmith \&~Rivlin~\cite{coppersmith&rivlin:poly}]
\label{thcopprivlin}
Every polynomial $p$ of degree $d \le n$ that has absolute value
\[
|p(i)|\leq 1 \mbox{ for all integers } i\in[0,n],
\]
satisfies
\[
|p(x)|< a e^{b d^2/n} \mbox{ for all real } x\in[0,n],
\]
where $a,b>0$ are universal constants (no explicit values for $a$ and $b$
are given in~\cite{coppersmith&rivlin:poly}).
\end{theorem}

The other two results concern the Chebyshev polynomials $T_d$, defined
as in~\cite{rivlin:chebyshev}:
\[
T_d(x)=\frac{1}{2}
\left(\left(x+\sqrt{x^2-1}\right)^d+\left(x-\sqrt{x^2-1}\right)^d\right).
\]
$T_d$ has degree $d$ and its absolute value $|T_d(x)|$ is bounded by 1 if
$x\in[-1,1]$.  On the interval $[1,\infty)$, $T_d$ exceeds all others
polynomials with those two properties (\cite[p.108]{rivlin:chebyshev}
and~\cite[Fact~2]{paturi:degree}):

\begin{theorem}
\label{thchebextremal}
If $q$ is a polynomial of degree $d$ such that $|q(x)|\leq 1$ for all
$x\in[-1,1]$ then $|q(x)|\leq|T_d(x)|$ for all $x\geq 1$.
\end{theorem}

Paturi~\cite[before Fact~2]{paturi:degree} proved

\begin{lemma}
[Paturi~\cite{paturi:degree}]
\label{lemchebbound}
$T_d(1+\mu)\leq e^{2d\sqrt{2\mu+\mu^2}}$ for all $\mu\geq 0$.
\end{lemma}

\begin{proof}
For $x=1+\mu$:
$\displaystyle T_d(x)\leq (x+\sqrt{x^2-1})^d=(1+\mu+\sqrt{2\mu+\mu^2})^d\leq
(1+2\sqrt{2\mu+\mu^2})^d \leq e^{2d\sqrt{2\mu+\mu^2}}$
(using that $1+z\leq e^z$ for all real $z$).
\end{proof}

\goodbreak
The following key lemma is the basis for all our direct product theorems:
\begin{lemma}\label{lem:polbound2}
Suppose $p$ is a degree-$D$ polynomial such that for some $\delta\geq 0$
\begin{quote}
$-\delta\leq p(i)\le\delta$ for all $i\in\{0,\ldots,k-1\}$,\\[1mm]
$p(k)=\sigma$,\\[1mm]
$p(i)\in[-\delta,1+\delta]$ for all $i\in\{0,\ldots,N\}$.
\end{quote}
Then for every integer $1\leq C<N-k$ and $\mu=2C/(N-k-C)$ we have
\[
\sigma
\leq
a\left(1+\delta+\frac{\delta(2N)^k}{(k-1)!}\right)\cdot
\exp\left(\frac{b(D-k)^2}{(N-k-C)}+2(D-k)\sqrt{2\mu+\mu^2}-k\ln(C/k)\right)
+\delta k 2^{k-1},
\]
where $a,b$ are the constants given by Theorem~\ref{thcopprivlin}.
\end{lemma}

\noindent
Before establishing this gruesome bound, let us reassure
the reader by noting that we will apply this lemma with $\delta$
negligibly small, $D=\alpha\sqrt{k N}$ for sufficiently small $\alpha$,
and $C=k e^{\gamma+1}$, giving
\[
\sigma\leq \exp\left((b\alpha^2+4\alpha e^{\gamma/2 + 1/2}-1-\gamma)k\right)\leq e^{-\gamma k}\leq 2^{-\gamma k}.
\]

\begin{proof}[Proof of Lemma~\ref{lem:polbound2}.]
Divide $p$ with remainder by $\prod_{j=0}^{k-1} (x-j)$ to obtain
\[
p(x) = q(x) \prod_{j=0}^{k-1} (x-j) + r(x),
\]
where $d=\deg(q)=D-k$ and $\deg(r) \le k-1$.
We know that $r(x)=p(x)\in[-\delta,\delta]$ for all $x \in\{0,\ldots,k-1\}$.
Decompose $r$ as a linear combination of polynomials $e_i$, where
$e_i(i) = 1$ and $e_i(x) = 0$ for $x \in\{0,\ldots,k-1\} - \{ i \}$:
\begin{eqnarray*}
r(x) &=& \sum_{i=0}^{k-1} p(i) e_i(x)
= \sum_{i=0}^{k-1} p(i) \prod_{\textstyle{j=0\atop j \ne i}}^{k-1} {x-j \over i-j}. \\
\noalign{\medskip\noindent
We bound the values of $r$ for all real $x\in[0,N]$ by
\medskip}
|r(x)| &\le& \sum_{i=0}^{k-1} {|p(i)| \over i! (k-1-i)!}
\prod_{\textstyle{j=0\atop j \ne i}}^{k-1} |x-j|\\
&\le& \frac{\delta}{(k-1)!}\sum_{i=0}^{k-1}{k-1\choose i}N^k
\le \frac{\delta(2N)^k}{(k-1)!}, \\
|r(k)| &\le& \delta k2^{k-1}. \\
\noalign{\medskip\noindent This implies the following about the values of
the polynomial $q$: \medskip}
|q(k)| &\geq& (\sigma-\delta k2^{k-1})/k!\\
|q(i)| &\leq& \frac{(i-k)!}{i!}\left(1+\delta+\frac{\delta(2N)^k}{(k-1)!}\right)
\quad \mbox{for $i\in\{k,\ldots,N\}$} \\
\noalign{\noindent In particular:}
|q(i)| &\leq& C^{-k}\left(1+\delta+\frac{\delta(2N)^k}{(k-1)!}\right)=A
\quad \mbox{for $i\in\{k+C,\ldots,N\}$} \\
\noalign{\medskip\noindent Theorem~\ref{thcopprivlin} implies that there are
constants $a,b>0$ such that\medskip}
|q(x)| &\leq& A\cdot a e^{b d^2/(N-k-C)}=B
\quad \mbox{for all real $x\in[k+C,N]$.}
\end{eqnarray*}
We now divide $q$ by $B$ to normalize it, and
rescale the interval $[k+C,N]$ to $[1,-1]$ to get a degree-$d$ polynomial $t$
satisfying
\begin{eqnarray*}
|t(x)| &\leq& 1 \quad \mbox{for all $x\in[-1,1]$} \\
t(1+\mu) &=& q(k)/B \quad \mbox{for $\mu=2C/(N-k-C)$} \\
\noalign{\medskip\noindent Since $t$ cannot grow faster than the degree-$d$
Chebyshev polynomial, we get\medskip}
t(1+\mu) &\leq& T_{d}(1+\mu)\leq e^{2d\sqrt{2\mu+\mu^2}}.
\end{eqnarray*}
Combining our upper and lower bounds on $t(1+\mu)$, we obtain
\[
\frac{(\sigma - \delta k 2^{k-1})/k!}
{C^{-k} \left( 1 + \delta + {\delta (2N)^k \over (k-1)!} \right)
a e^{b d^2/(N-k-C)}}
\le e^{2 d \sqrt{2 \mu + \mu^2}}.
\]
Rearranging gives the bound.
\end{proof}

\subsection{Consequences for quantum algorithms}

The previous result about polynomials implies a strong tradeoff
between queries and success probability for quantum algorithms
that have to find $k$ ones in an $N$-bit input.
A \emph{$k$-threshold algorithm with success probability $\sigma$} is
an algorithm on $N$-bit input $x$, that outputs 0 with certainty
if $|x|<k$, and outputs 1 with probability at least $\sigma$ if $|x|=k$.

\begin{theorem}\label{th:search}
For every $\gamma > 0$, there exists an $\alpha > 0$ such
that every quantum $k$-threshold algorithm with $T\leq \alpha\sqrt{k N}$ queries
has success probability $\sigma\leq 2^{-\gamma k}$.
\end{theorem}

\begin{proof}
Fix $\gamma>0$ and consider a $T$-query $k$-threshold algorithm.
By~\cite{bbcmw:polynomialsj}, its acceptance probability is an
$N$-variate polynomial of degree $D\leq 2T\leq 2\alpha\sqrt{k N}$
and can be symmetrized to a single-variate polynomial $p$ with the properties
\begin{quote}
$p(i) = 0$ if $i\in\{0,\ldots,k-1\}$\\
$p(k)\geq \sigma$\\
$p(i) \in [0,1]$ for all $i\in\{0,\ldots,N\}$
\end{quote}
Choosing $\alpha>0$ sufficiently small and $\delta=0$,
the result follows from Lemma~\ref{lem:polbound2}.
\end{proof}

This implies a strong direct product theorem for $k$ instances of the $n$-bit search problem:

\begin{theorem}[SQDPT for Search]\label{th:ind-search}
For every $\gamma > 0$, there exists an $\alpha > 0$ such
that every quantum algorithm for $\vectorres{\Search_n}k$ with $T\leq \alpha k\sqrt{n}$
queries has success probability $\sigma \leq 2^{-\gamma k}$.
\end{theorem}

\begin{proof}
Set $N=k n$, fix a $\gamma>0$ and a $T$-query algorithm $A$ for $\vectorres{\Search_n}k$
with success probability $\sigma$.
Now consider the following algorithm that acts on an $N$-bit input $x$:
\begin{enumerate}
\item Apply a random permutation $\pi$ to $x$.
\item Run $A$ on $\pi(x)$.
\item Query each of the $k$ positions that $A$ outputs,
and return 1 iff at least $k/2$ of those bits are 1.
\end{enumerate}
This uses $T+k$ queries. We will show that it is a $k/2$-threshold algorithm.
First, if $|x|<k/2$, it always outputs 0.
Second, consider the case $|x|=k/2$. The probability that $\pi$
puts all $k/2$ ones in distinct $n$-bit blocks is
\[
{N \over N} \cdot {N-n \over N-1} \cdots {N-\frac k2 n \over N-\frac k2}
\ge \left( N - \frac k2 n \over N \right)^{k/2}
= 2^{-k/2}.
\]
Hence our algorithm outputs 1 with probability at least $\sigma 2^{-k/2}$.
Choosing $\alpha$ sufficiently small, the previous theorem implies
$\sigma 2^{-k/2}\leq 2^{-(\gamma+1/2)k}$, hence $\sigma \leq 2^{-\gamma k}$.
\end{proof}

Our bounds are quite precise for $\alpha\ll 1$.
We can choose $\gamma=2\ln(1/\alpha)-\Oh{1}$ and ignore some
lower-order terms to get roughly $\sigma\leq\alpha^{2k}$.
On the other hand, it is known that Grover's search algorithm with
$\alpha\sqrt{n}$ queries on an $n$-bit input has success probability
roughly $\alpha^2$~\cite{bbht:bounds}.
Doing such a search on all $k$ instances gives overall
success probability $\alpha^{2k}$.


\begin{theorem}[SQDPT for \OR]\label{th:vector-ors}
There exist $\alpha,\gamma>0$ such that every quantum algorithm
for $\vectorres{\OR_n}k$ with $T\leq \alpha k\sqrt{n}$ queries has
success probability $\sigma \leq 2^{-\gamma k}$.
\end{theorem}

\begin{proof}
An algorithm $A$ for $\vectorres{\OR_n}k$ with success probability $\sigma$
can be used to build an algorithm $A'$ for $\vectorres{\Search_n}k$
with slightly worse success probability:
\begin{enumerate}
\item Run $A$ on the original input and remember which blocks contain a 1.
\item Run simultaneously (at most $k$) binary searches on the nonzero blocks.
Iterate this $s=2\log(1/\alpha)$ times. Each iteration is computed by running $A$ on
the parts of the blocks that are known to contain a 1, halving the remaining
instance size each time.
\item Run the exact version of Grover's algorithm on each of the
remaining parts of the instances to look for a one there
(each remaining part has size $n/2^s$).
\end{enumerate}
This new algorithm $A'$ uses
$(s+1)T+\frac{\pi}{4}k\sqrt{n/2^s}=\Oh{\alpha\log(1/\alpha)k\sqrt{n}}$
queries. With probability at least $\sigma^{s+1}$, $A$ succeeds in all iterations,
in which case $A'$ solves $\vectorres{\Search_n}k$.
By the previous theorem, for every $\gamma'>0$ of our choice we can
choose $\alpha>0$ such that
$$
\sigma^{s+1}\leq 2^{-\gamma' k},
$$
which implies the theorem with $\gamma=\gamma'/(s+1)$.
\end{proof}

Choosing our parameters carefully, we can actually show
that for every $\gamma<1$ there is an $\alpha>0$ such that $\alpha k\sqrt{n}$
queries give success probability $\sigma\leq 2^{-\gamma k}$.
Clearly, $\sigma=2^{-k}$ is achievable without any queries by random guessing.

\subsection{A bound for all functions}

As in Section~\ref{sec:all-func}, we can extend the strong direct product
theorem for \OR\ to a slightly weaker theorem for all total functions.
Block sensitivity is closely related to bounded-error quantum query complexity:

\begin{theorem}[\cite{bbcmw:polynomialsj}]
$Q_2(f)=\Om{\sqrt{\bs(f)}}$ for all $f$,
$D(f)\le\bs(f)^3$ for all total Boolean $f$.
\end{theorem}

By embedding an \OR\ of size $\bs(f)$ in $f$, we obtain

\begin{theorem}
There exist $\alpha,\gamma > 0$ such that for every $f$,
every quantum algorithm for $\vectorres{f}k$
with $T\leq \alpha k\sqrt{\bs(f)}$ queries
has success probability $\sigma \leq 2^{-\gamma k}$.
\end{theorem}

This is close to optimal whenever $Q_2(f)=\Th{\sqrt{\bs(f)}}$.
For total functions, the gap between $Q_2(f)$ and $\sqrt{\bs(f)}$
is no more than a 6th power, hence

\begin{corollary}
There exist $\alpha,\gamma > 0$ such that for every total Boolean $f$,
every quantum algorithm for $\vectorres{f}k$ with
$T\leq \alpha k Q_2(f)^{1/6}$ queries has success
probability $\sigma \leq 2^{-\gamma k}$.
\end{corollary}

\section{Strong Direct Product Theorem for Quantum Communication}\label{secquacomdpt}

In this section we establish a strong direct product theorem
for quantum communication complexity, specifically for protocols
that compute $k$ independent instances of the Disjointness problem.
Our proof relies crucially on the beautiful technique that Razborov
introduced to establish a lower bound on the quantum communication
complexity of (one instance of) Disjointness~\cite{razborov:qdisj}.
It allows us to translate a quantum communication
protocol to a single-variate polynomial that represents,
roughly speaking, the protocol's acceptance probability as
a function of the size of the intersection of $x$ and $y$.
Once we have this polynomial, the results from Section~\ref{ssecpolynomial}
suffice to establish a strong direct product theorem.

\subsection{Razborov's technique}

Razborov's technique relies on the following linear algebraic notions.
The \emph{operator norm} $\norm{A}$ of a matrix $A$ is its largest singular value $\sigma_1$.
The \emph{trace inner product} between $A$ and $B$ is $\inpc{A}{B}=\Tr(A^*B)$.
The \emph{trace norm} is $\norm{A}_{tr}=\max\{|\inpc{A}{B}|:\norm{B}=1\}=\sum_i\sigma_i$.
The \emph{Frobenius norm} is $\norm{A}_F=\sqrt{\sum_{i j}|A_{i j}|^2}=\sqrt{\sum_i\sigma_i^2}$.
The following lemma is implicit in Razborov's paper.

\begin{lemma}\label{lemrazborov}
Consider a $Q$-qubit quantum communication protocol on $N$-bit inputs $x$ and $y$,
with acceptance probabilities denoted by $P(x,y)$.
Define $P(i)={\rm E}_{|x|=|y|=N/4, |x\wedge y|=i|}[P(x,y)]$,
where the expectation is taken uniformly over all $x,y$
that each have weight $N/4$ and that have intersection $i$.
For every $d\leq N/4$ there exists a degree-$d$ polynomial $q$ such
that $|P(i)-q(i)|\leq 2^{-d/4+2Q}$ for all $i\in\{0,\ldots,N/8\}$.
\end{lemma}

\begin{proof}
We only consider the ${\cal N}={N\choose N/4}$ strings of weight $N/4$.
Let $P$ denote the ${\cal N}\times {\cal N}$ matrix of the
acceptance probabilities on these inputs.
We know from Yao and Kremer~\cite{yao:qcircuit,kremer:thesis}
that we can decompose $P$ as a matrix product $P=A B$,
where $A$ is an ${\cal N}\times 2^{2Q-2}$ matrix with each
entry at most 1 in absolute value, and similarly for $B$.
Note that $\norm{A}_F,\norm{B}_F\leq\sqrt{{\cal N}2^{2Q-2}}$.
Using H\"older's inequality we have:
$$
\norm{P}_{tr}\leq\norm{A}_F\cdot\norm{B}_F\leq {\cal N}2^{2Q-2}.
$$
Let $\mu_i$ denote the ${\cal N}\times {\cal N}$ matrix corresponding
to the uniform probability distribution on $\{(x,y) : |x\wedge y|=i\}$.
These ``combinatorial matrices'' have been well studied~\cite{knuth:cm}.
Note that $\inpc{P}{\mu_i}$ is the expected acceptance
probability $P(i)$ of the protocol under that distribution.
One can show that the different $\mu_i$ commute, so they have
the same eigenspaces $E_0,\ldots,E_{N/4}$ and can be simultaneously
diagonalized by some orthogonal matrix $U$.
For $t\in\{0,\ldots,N/4\}$, let $(U P U^T)_t$ denote the block of
$U P U^T$ corresponding to $E_t$, and $a_t=\Tr((U P U^T)_t)$ be its trace.
Then we have
$$
\sum_{t=0}^{N/4} |a_t|\leq\sum_{j=1}^{{\cal N}} \left|(U P U^T)_{jj}\right|
\leq\norm{U P U^T}_{tr}=\norm{P}_{tr}\leq {\cal N}2^{2Q-2},
$$
where the second inequality is a property of the trace norm.

Let $\lambda_{it}$ be the eigenvalue of $\mu_i$ in eigenspace $E_t$.
It is known~\cite[Section~5.3]{razborov:qdisj} that $\lambda_{it}$
is a degree-$t$ polynomial in $i$,
and that $|\lambda_{it}|\leq 2^{-t/4}/{\cal N}$ for $i\leq N/8$
(the factor $1/4$ in the exponent is implicit in Razborov's paper).
Consider the high-degree polynomial $p$ defined by
$$
p(i)=\sum_{t=0}^{N/4} a_t\lambda_{it}.
$$
This satisfies
$$
p(i)=\sum_{t=0}^{N/4} \Tr((U P U^T)_t)\lambda_{it}=\inpc{U P U^T}{U\mu_i U^T}=\inpc{P}{\mu_i}=P(i).
$$
Let $q$ be the degree-$d$ polynomial obtained by removing the high-degree parts of $p$:
$$
q(i)=\sum_{t=0}^d a_t\lambda_{it}.
$$
Then $P$ and $q$ are close on all integers $i$ between 0 and $N/8$:
$$
|P(i)-q(i)|=|p(i)-q(i)|=\left|\sum_{t=d+1}^{N/4} a_t\lambda_{it}\right|\leq
\frac{2^{-d/4}}{{\cal N}}\sum_{t=0}^{N/4}|a_t|\leq 2^{-d/4+2Q}.
$$
\vspace*{-1em}
\end{proof}

\subsection{Consequences for quantum protocols}

Combining Razborov's technique with our polynomial bounds we can prove

\begin{theorem}[SQDPT for Disjointness]\label{th:vector-disj}
There exist $\alpha,\gamma > 0$ such that every quantum protocol for
$\vectorres{\DISJ_n}k$ with $Q\leq \alpha k\sqrt{n}$ qubits of
communication has success probability $p\leq 2^{-\gamma k}$.
\end{theorem}

\begin{proof}[Proof (sketch).]
By doing the same trick with $s=2\log(1/\alpha)$ rounds of
binary search as for Theorem~\ref{th:vector-ors},
we can tweak a protocol for $\vectorres{\DISJ_n}k$ to a protocol
that satisfies, with $P(i)$ defined as in Lemma~\ref{lemrazborov},
$N=k n$ and $\sigma=p^{s+1}$:
\begin{quote}
$P(i) = 0$ if $i\in\{0,\ldots,k-1\}$\\
$P(k)\geq \sigma$\\
$P(i) \in [0,1]$ for all $i\in\{0,\ldots,N\}$
\end{quote}
(a subtlety: instead of exact Grover we use an exact version of the
$\Oh{\sqrt{n}}$-qubit Disjointness protocol of~\cite{aaronson&ambainis:search};
the~\cite{BuhrmanCleveWigderson98}-protocol would lose a $\log n$-factor).
Lemma~\ref{lemrazborov}, using $d=12 Q$, then gives a degree-$d$
polynomial $q$ that differs from $P$ by at most $\delta \le 2^{-Q}$
on all $i\in\{0,\ldots,N/8\}$.
This $\delta$ is sufficiently small to apply Lemma~\ref{lem:polbound2},
which in turn upper bounds $\sigma$ and hence $p$.
\end{proof}

\noindent
This technique also gives strong direct product theorems for
symmetric predicates other than $\DISJ_n$.

\section{Time-Space Tradeoff for Quantum Sorting}

We will now use our strong direct product theorem to get
near-optimal time-space tradeoffs for quantum circuits for
sorting. This follows Klauck~\cite{klauck:qsorting}, who described
an upper bound $T^2S=\Oh{(N\log N)^3}$ and a lower bound
$TS=\Om{N^{3/2}}$. In our model, the numbers $a_1,\ldots,a_N$ that
we want to sort can be accessed by means of queries, and the
number of queries lower bounds the actual time taken by the
circuit.
The circuit has $N$ output gates and in the course of its computation
outputs the $N$ numbers in sorted (say, descending) order, with
success probability at least $2/3$.

\begin{theorem}
Every bounded-error quantum circuit for sorting $N$ numbers that uses
$T$ queries and $S$ qubits of workspace satisfies $T^2S=\Om{N^3}$.
\end{theorem}

\begin{proof}
We ``slice'' the circuit along the time-axis into
$L=T/\alpha\sqrt{SN}$ slices, each containing
$T/L=\alpha\sqrt{SN}$ queries.  Each such slice has a number of
output gates. Consider any slice. Suppose it contains output gates
$i,i+1,\ldots, i+k-1$, for $i\leq N/2$, so it is supposed to
output the $i$-th up to $i+k-1$-th largest elements of its input.
We want to show that $k=\Oh S$. If $k\leq S$ then we are done, so
assume $k>S$. We can use the slice as a $k$-threshold algorithm on
$N/2$ bits, as follows. For an $N/2$-bit input $x$, construct a
sorting input by taking $i-1$ copies of the number $2$, the $N/2$
bits in $x$, and $N/2-i+1$ copies of the number $0$,
and append their position behind the numbers.

Consider the behavior of the sorting circuit on this input. The
first part of the circuit has to output the $i-1$ largest numbers,
which all start with 2.  We condition on the event that the
circuit succeeds in this. It then passes on an $S$-qubit state
(possibly mixed) as the starting state of the particular slice we
are considering. This slice then outputs the $k$ largest numbers
in $x$ with probability at least $2/3$. Now, consider an algorithm
that runs just this slice, starting with the completely mixed
state on $S$-qubits, and that outputs 1 if it finds $k$ numbers
starting with 1, and outputs 0 otherwise. If $|x|<k$ this new
algorithm always outputs 0 (note that it can verify finding a 1
since its position is appended), but if $|x|=k$ then it outputs 1
with probability at least $\sigma \geq \frac{2}{3}\cdot 2^{-S}$,
because the completely mixed state has ``overlap'' $2^{-S}$ with
the ``good'' $S$-qubit state that would have been the starting
state of the slice in the run of the sorting circuit. On the other
hand, the slice has only $\alpha\sqrt{SN}<\alpha\sqrt{k N}$
queries, so by choosing $\alpha$ sufficiently small,
Theorem~\ref{th:search} implies $\sigma \leq 2^{-\Om k}$.
Combining our upper and lower bounds on $\sigma$ gives $k=\Oh S$.
Thus we need $L=\Om{N/S}$ slices, so
$T=L\alpha\sqrt{SN}=\Om{N^{3/2}/\sqrt{S}}$.
\end{proof}

As mentioned, our tradeoff is achievable up to polylog
factors~\cite{klauck:qsorting}. Interestingly, the near-optimal
algorithm uses only a polylogarithmic number of qubits and
otherwise just classical memory. For simplicity we have shown the
lower bound for the case when the outputs have to be made in their
natural ordering only, but we can show the same lower bound for
any ordering of the outputs that does not depend on the input
using a slightly different proof.

\section{Time-Space Tradeoffs for Boolean Matrix Products}

First we show a lower bound on the time-space tradeoff for Boolean
matrix-vector multiplication on \emph{classical} machines.

\begin{theorem} \label{th:matrixvector}
There is a matrix $A$ such that every classical bounded-error circuit that
computes the Boolean matrix-vector product $Ab$ with $T$ queries
and space $S=\oo{N/\log N}$ satisfies $TS=\Om{N^2}$.
\end{theorem}

The bound is tight if $T$ measures queries to the input.
\medskip

\begin{proof}
Fix $k=\Oh S$ large enough.
First we have to find a hard matrix $A$. We pick $A$ randomly
by setting $N/(2k)$ random positions in each row to 1. We want to
show that with positive probability for all sets of $k$ rows
$A[i_1],\ldots,A[i_k]$ many of the rows $A[i_j]$ contain at least
$N/(6k)$ ones that are not ones in any of the $k-1$ other rows.

This probability can be bounded as follows. We will treat the rows
as subsets of $\{1,\ldots, N\}$. A row $A[j]$ is
called \emph{bad} with respect to $k-1$ other rows
$A[i_1],\ldots,A[i_{k-1}]$, if $|A[j]-\cup_\ell A[i_\ell]|\le N/(6k)$.
For fixed $i_1,\ldots,i_{k-1}$, the probability that some $A[j]$ is
bad with respect to the $k-1$ other rows is at most
$e^{-\Om{N/k}}$ by the Chernoff bound and the fact that $k$
rows can together contain at most $N/2$ elements.
Since $k=\oo{N/\log N}$ we may assume this probability is at most
$1/N^{10}$.

Now fix any set $I=\{i_1,\ldots,i_k\}$. The probability that for
$j\in I$ it holds that $A[j]$ is bad with respect to the other
rows is at most $1/N^{10}$, and this also holds, if we condition
on the event that some other rows are bad, since this condition
makes it only less probable that another row is also bad.
So for any fixed $J\subset I$ of size $k/2$ the probability that
all rows in $J$ are bad is at most $N^{-5k}$, and the probability
that there exists such $J$ is at most
\[{k\choose k/2} N^{-5k}.\]
Furthermore the probability that there is a set $I$ of $k$ rows
for which $k/2$ are bad is at most
\[{N\choose k} {k\choose k/2} N^{-5k}<1.\]
So there is an $A$ as required and we may fix one.

Now suppose we are given a circuit with space $S$ that computes
the Boolean product between the rows of $A$ and $b$ in some order.
We again proceed by ``slicing'' the circuit into $L=T/\alpha N$
slices, each containing $T/L=\alpha N$ queries.  Each such
slice has a number of output gates. Consider any slice. Suppose it
contains output gates $i_1<\ldots<i_k\leq N/2$, so it is supposed
to output $\vee_{\ell=1}^N\left(A[i_j,\ell]\wedge b_\ell\right)$
for all $i_j$ with $1\le j\le k$.

Such a slice starts on a classical value of the ``memory'' of the
circuit, which is in general a probability distribution on $S$
bits (if the circuit is randomized). We replace this probability
distribution by the uniform distribution on the possible values of
$S$ bits. If the original circuit succeeds in computing the
function correctly with probability at least $1/2$, then so does
the circuit slice with its outputs, and replacing the initial
value of the memory by a uniformly random one decreases the
success probability to no less than $(1/2)\cdot 1/2^S$.

If we now show that any classical circuit with $\alpha N$ queries
that produces the outputs $i_1,\ldots, i_k$ can succeed only with
exponentially small probability in $k$, we get that $k=\Oh S$, and
hence $(T/\alpha N) \cdot \Oh S \ge N$, which gives the claimed
lower bound for the time-space tradeoff.

Each set of $k$ outputs corresponds to $k$ rows of $A$,
which contain $N/(2k)$ ones each. Thanks to the construction of
$A$ there are $k/2$ rows among these, such that $N/(6k)$ of the
ones in each such row are in position where none of the other
contains a one. So we get $k/2$ sets of $N/(6k)$ positions that
are unique to each of the $k/2$ rows. The inputs for $b$ will be
restricted to contain ones only at these positions, and so the
algorithm naturally has to solve $k/2$ independent \OR\ problems
on $n=N/(6k)$ bits each.  By Theorem~\ref{th:classor}, this is only
possible with $\alpha N$ queries if the success probability
is exponentially small in $k$.
\end{proof}

An absolutely analogous construction can be done in the quantum case.
Using circuit slices of length $\alpha\sqrt{NS}$ we can prove:

\begin{theorem}
There is a matrix $A$ such that every quantum bounded-error circuit that
computes the Boolean matrix-vector product $Ab$ with $T$ queries
and space $S=\oo{N/\log N}$ satisfies $T^2 S=\Om{N^{3}}$.
\end{theorem}

Note that this is tight within a logarithmic factor (needed to
improve the success probability of Grover search).


\begin{theorem}\label{th:matrixmatrix}
Every classical bounded-error circuit that computes the Boolean matrix product
$A B$ with $T$ queries and space $S$ satisfies $TS=\Om{N^{3}}$.
\end{theorem}

While this is near-optimal for small $S$, it is probably not tight
for large $S$, a likely tight tradeoff being $T^2S=\Om{N^6}$.
It is also no improvement compared to Abrahamson's average-case
bounds~\cite{abrahamson:booleantrade}.

\medskip

\begin{proof}
Suppose that $S=o(N)$, otherwise the bound
is trivial, since time $N^2$ is always needed. We can proceed
similar to the proof of Theorem~\ref{th:matrixvector}. We slice
the circuit so that each slice has only $\alpha N$ queries.
Suppose a slice makes $k$ outputs. We are going to restrict the
inputs to get a direct product problem with $k$ instances of size
$N/k$ each, hence a slice with $\alpha N$ queries has
exponentially small success probability in $k$ and can thus
produce only $\Oh S$ outputs. Since the overall number of outputs
is $N^2$ we get the tradeoff $TS=\Om{N^3}$.

Suppose a circuit slice makes $k$ outputs, where an output
labeled $(i,j)$ needs to produce the vector product of the $i$th
row $A[i]$ of $A$ and the $j$th column $B[j]$ of $B$. We may
partition the set $\{1,\ldots,N\}$ into $k$ mutually disjoint
subsets $U(i,j)$ of size $N/k$, each associated to an output
$(i,j)$.

Assume that there are $\ell$ outputs $(i,j_1),\ldots, (i,j_\ell)$
involving $A[i]$. Each such output is associated to a subset
$U(i,j_t)$, and we set $A[i]$ to zero on all positions that are
not in any of these subsets, and to one on all positions that are
in one of these. When there are $\ell$ outputs $(i_1,j),\ldots,
(i_\ell,j)$ involving $B[j]$, we set $B[j]$ to zero on all
positions that are not in any of the corresponding subsets, and
allow the inputs to be arbitrary on the other positions.

If the circuit computes on these restricted inputs, it actually
has to compute $k$ instances of $\OR$ of size $n=N/k$ in $B$, for
it is true that $A[i]$ and $B[j]$ contain a single block of size $N/k$
in which $A[i]$ contains only ones, and $B[j]$ ``free'' input bits,
if and only if $(i,j)$ is one of the $k$ outputs. Hence the strong
direct product theorem is applicable.
\end{proof}

The application to the quantum case is analogous.

\begin{theorem}
Every quantum bounded-error circuit that computes the Boolean matrix product
$A B$ with $T$ queries and space $S$ satisfies $T^2 S=\Om{N^{5}}$.
\end{theorem}

\noindent
If $S=\Oh{\log N}$, then $N^2$ applications of Grover can compute $A B$
with $T=\Oh{N^{2.5}\log N}$. Hence our tradeoff
is near-optimal for small $S$. We do not know whether it is
optimal for large $S$.

\section{Quantum Communication-Space Tradeoffs for Matrix Products}

In this section we use the strong direct product result for quantum
communication (Theorem~\ref{th:vector-disj}) to prove
communication-space tradeoffs.
We later show that these are close to optimal.

\begin{theorem}
Every quantum bounded-error protocol in which Alice and Bob have bounded space
$S$ and that computes the Boolean matrix-vector product, satisfies
$C^2 S=\Om{N^{3}}$.
\end{theorem}

\begin{proof}
In a protocol, Alice receives a matrix $A$, and Bob a vector $b$ as
inputs. Given a circuit that multiplies these with communication
$C$ and space $S$, we again proceed to slice it. This time,
however, a slice contains a limited amount of communication.
Recall that in communicating quantum circuits the communication
corresponds to wires carrying qubits that cross between Alice's
and Bob's circuits. Hence we may cut the circuit after
$\alpha\sqrt{NS}$ qubits have been communicated and so on. Overall
there are $C/\alpha\sqrt{NS}$ circuit slices. Each starts with an
initial state that may be replaced by the completely mixed
state at the cost of decreasing the success probability to
$(1/2)\cdot 1/2^S$. We want to employ the direct product theorem for
quantum communication complexity to show that a protocol with the
given communication has success probability at most
exponentially small in the number of outputs it produces,
and so a slice can produce at most $\Oh{S}$
outputs. Combining these bounds with the fact that $N$
outputs have to be produced gives the tradeoff.

To use the direct product theorem we restrict the inputs in the
following way: Suppose a protocol makes $k$ outputs. We partition
the vector $b$ into $k$ blocks of size $N/k$, and each block is
assigned to one of the $k$ rows of $A$ for which an output is
made. This row is made to contain zeroes outside of the positions
belonging to its block, and hence we arrive at a problem where
Disjointness has to be computed on $k$ instances of size $N/k$.
With communication $\alpha \sqrt{k N}$, the success probability must
be exponentially small in $k$ due to Theorem~\ref{th:vector-disj}.
Hence $k=\Oh S$ is an upper bound on the number of outputs produced.
\end{proof}

\begin{theorem}
Every quantum bounded-error protocol in which Alice and Bob have bounded space
$S$ and that computes the Boolean matrix product, satisfies
$C^2 S=\Om{N^{5}}$.
\end{theorem}

\begin{proof}
The proof uses the same slicing approach as in the other tradeoff
results. Note that we can assume that $S=\oo{N}$, since otherwise
the bound is trivial. Each slice contains communication $\alpha
\sqrt{NS}$, and as before a direct product result showing that $k$
outputs can be computed only with success probability
exponentially small in $k$ leads to the conclusion that a slice
can only compute $\Oh S$ outputs. Therefore $(C/\alpha\sqrt{NS})\cdot \Oh S \ge
N^2$, and we are done.

Consider a protocol with
$\alpha\sqrt{NS}$ qubits of communication. We
partition the universe $\{1,\ldots,N\}$ of the Disjointness
problems to be computed into $k$ mutually disjoint subsets
$U(i,j)$ of size $N/k$, each associated to an output $(i,j)$,
which in turn corresponds to a row/column pair $A[i]$, $B[j]$ in
the input matrices $A$ and $B$. Assume that there are $\ell$
outputs $(i,j_1),\ldots, (i,j_\ell)$ involving $A[i]$. Each output
is associated to a subset of the universe $U(i,j_t)$, and we set
$A[i]$ to zero on all positions that are not in one of these
subsets. Then we proceed analogously with the columns of $B$.

If the protocol computes on these restricted inputs, it
has to solve $k$ instances of Disjointness of size $n=N/k$ each,
since $A[i]$ and $B[j]$ contain a single block of size
$N/k$ in which both are not set to 0 if and only if $(i,j)$ is one
of the $k$ outputs. Hence Theorem~\ref{th:vector-disj} is applicable.
\end{proof}

We now want to show that these tradeoffs are not too far from
optimal.

\begin{theorem}
There is a quantum bounded-error protocol with space $S$ that computes the
Boolean product between a matrix and a vector within communication
$C=\OO{(N^{3/2}\log^{2} N)/\sqrt S}$.

There is a quantum bounded-error protocol with space $S$ that computes the
Boolean product between two matrices within communication
$C=\OO{(N^{5/2}\log^{2} N)/\sqrt S}$.
\end{theorem}

\begin{proof}
We begin by showing a protocol for the following scenario:
Alice gets $S$ $N$-bit vectors $x_1,\ldots,x_S$, Bob gets
an $N$-vector vector $y$, and they want to compute the $S$ Boolean
inner products between these vectors. The protocol uses space $\Oh S$.

In the following, we interpret Boolean vectors as sets. The main
idea is that Alice can use the union $z$ of the $x_i$ and then
Alice and Bob can find an element in the intersection of $z$ and
$y$ using the protocol for the Disjointness problem described in
\cite{BuhrmanCleveWigderson98}. Alice then marks all $x_i$ that
contain this element and removes them from $z$.

A problem with this approach is that Alice cannot store $z$
explicitly, since it might contain much more than $S$ elements.
Alice may, however, store the indices of those sets $x_i$ for
which an element in the intersection of $x_i$ and $y$ has already
been found, in an array of length $S$. This array and the input
given as an oracle work as an implicit representation of $z$.

Now suppose at some point during the protocol the intersection of
$z$ and $y$ has size $k$. Then Alice and Bob can find one element
in this intersection within $\OO{\sqrt{N/k}}$ rounds of
communication in which $\Oh{\log N}$ qubits are exchanged each.
Furthermore in $\OO{\sqrt{N k}}$ rounds all elements in the
intersection can be found. So if $k\le S$, then all elements are found
within communication $\OO{\sqrt{NS}\log N}$ and the problem can be
solved completely. On the other hand, if $k\ge S$, finding one
element costs $\OO{\sqrt{N/S}\log N}$, but finding such an element
removes at least one $x_i$ from $z$, and hence this has to be done
at most $S$ times, giving the same overall communication bound.

It is not hard to see that this process can be implemented with
space $\Oh S$. The protocol from \cite{BuhrmanCleveWigderson98} is
a distributed Grover search that itself uses only space $\OO{\log
N}$. Bob can work as in this protocol. For each search, Alice has
to start with a superposition over all indices in $z$. This
superposition can be computed from her oracle and her array. In
each step she has to apply the Grover iteration. This can also be
implemented from these two resources.

To get a protocol for matrix-vector product, the above procedure is
repeated $N/S$ times, hence the communication is $\OO{(N/S)\cdot
\sqrt{NS}\log^{2}N}$, where one logarithmic factor stems from
improving success probability to $1/\poly(N)$.

For the product of two matrices, the matrix-vector protocol may
be repeated $N$ times.
\end{proof}

\noindent
These near-optimal protocols use only
$\OO{\log N}$ ``real'' qubits, and otherwise just classical memory.

\section{Open Problems}

We mention some open problems. The
first is to determine tight time-space tradeoffs for
Boolean matrix product on both classical and quantum computers.
Second, regarding communication-space tradeoffs for Boolean
matrix-vector and matrix product, we did not prove any classical
bounds that were better than our quantum bounds.
Klauck~\cite{klauck:tradeoffs} recently proved classical tradeoffs
$CS^2=\Om{N^3}$ and $CS^2=\Om{N^2}$ for Boolean matrix product and
matrix-vector product, respectively, by means of a \emph{weak}
direct product theorem for Disjointness. A classical \emph{strong}
direct product theorem for Disjointness would imply optimal tradeoffs,
but we do not know how to prove this at the moment.
Finally, it would be interesting to get any lower bounds
on time-space or communication-space tradeoffs for decision
problems in the quantum case, for example for
Element Distinctness~\cite{betal:distinctness,ambainis:ed}
or the verification of matrix multiplication~\cite{bs:matrix}.

\subsection*{Acknowledgments}
We thank Scott Aaronson for email discussions about the evolving
results in his~\cite{aaronson:advicecomm} that motivated some of our proofs,
and Harry Buhrman for useful discussions.

\bibliographystyle{alpha}

\newcommand{\etalchar}[1]{$^{#1}$}

\end{document}